\documentclass
{aa}

\newcommand{\ssty}{\scriptscriptstyle}
\newcommand{\tsty}{\textstyle}
\newcommand{\be}{\begin{equation}}
\newcommand{\ee}{\end{equation}}

\newcommand{\obs}[1]{[#1]_{{\tsty {\ssty \mathrm{obs}}}}}

\newcommand{\drm}{\mathrm{d}}
\newcommand{\mg}{\mathcal{M}_g}
\usepackage{txfonts}
\usepackage{graphicx}

\begin{document}
\title{Galaxy cosmological mass function}
\author{Amanda R.\ Lopes, \inst{1}\thanks{amanda05@astro.ufrj.br}
~       Alvaro Iribarrem, \inst{1}
~       Marcelo B.\ Ribeiro~~\inst{2}
\and    William R.\ Stoeger~~\inst{3}\thanks{In memoriam (1943-2014)}}
\institute{Observat\'orio do Valongo, Universidade Federal do Rio de
           Janeiro, Brazil
\and
Instituto de F\'{\i}sica, Universidade Federal do Rio de Janeiro, Brazil
\and
Vatican Observatory Research Group, Steward Observatory, University of
Arizona, USA}
\date{}
\abstract{}
{This paper studies the galaxy cosmological mass function (GCMF)
in a semi-empirical relativistic approach that uses 
observational data provided by recent galaxy redshift
surveys.}
{Starting from a previously presented relation 
between the mass-to-light ratio, the selection function obtained
from the luminosity function (LF) data and the luminosity density,
the average luminosity $L$, and the average galactic mass $\mg$ were 
computed in terms of the redshift. $\mg$ was also alternatively
estimated by means of a method that uses the galaxy stellar mass
function (GSMF). Comparison of these two forms of deriving the
average galactic mass allowed us to infer a possible bias
introduced by the selection criteria of the survey. We used the
FORS Deep Field galaxy survey sample of 5558 galaxies in the
redshift range $0.5 < z < 5.0$ and its LF Schechter parameters in
the B-band, as well as this sample's stellar mass-to-light ratio
and its GSMF data.}
{Assuming ${\mathcal{M}_{g_0}} \approx 10^{11} \mathcal{M}_\odot$
as the local value of the average galactic mass, the LF approach
results in $L_{B} \propto (1+z)^{(2.40 \pm 0.03)}$ and $\mg
\propto (1+z)^{(1.1\pm0.2)}$. However, using the GSMF
results to calculate the average galactic mass produces $\mg
\propto (1+z)^{(-0.58 \pm 0.22)}$. We chose the latter result
because it is less biased. We then obtained the theoretical quantities
of interest, such as the differential number counts, 
to finally calculate the GCMF, which can be fitted by
a Schechter function, but whose fitted parameter values are
different from the values found in the literature for the GSMF.} 
{This GCMF behavior follows the theoretical predictions
from the cold dark matter models in which the less massive
objects form first, followed later by more massive ones.
In the range $0.5 < z < 2.0$ the GCMF has a strong variation
that can be interpreted as a higher rate of galaxy mergers or
as a strong evolution in the star formation history
of these galaxies.}
\titlerunning{The Galaxy Cosmological Mass Function}
\authorrunning{Lopes et al.}
\maketitle

\section{Introduction}
The \textit{galaxy cosmological mass function} (GCMF) $\zeta$ is a
quantity defined in the framework of relativistic cosmology that 
measures the distribution of galactic masses in a given volume
within a certain redshift range of an evolving universe defined by a 
spacetime geometry. This is, nevertheless, a generic definition that
can be turned into an operational one by following the approach 
advanced by Ribeiro \& Stoeger (2003; hereafter RS03), which connects
the mass-energy density given by the right-hand side of Einstein field
equations, and the associated theoretically derived galaxy number counts,
with the astronomically determined luminosity function and mass-to-light
ratio. In this way, the GCMF contains information about the number
density evolution of all galaxies at a certain $z$, as well as their
average mass $\mg(z)$ in that redshift. Therefore, $\zeta \drm \mg$
provides the number density of galaxies with mass in the range $\mg$,
$\mg + \drm \mg$. Since $\mg(z)$ is the average mass at a specific
redshift value, the quantity $\zeta \drm \mg$ is given in the redshift
range $z$, $z+\drm z$.

In the astrophysical literature one can a find a quantity bearing
similarities to the GCMF, the \textit{galaxy stellar mass function}
(GSMF), which describes the number density of galaxies per logarithmic
stellar mass interval. It can be computed using the galaxy stellar masses
obtained from galactic luminosities (e.g., Larson \& Tinsley 1978;
Jablonka \& Arimoto 1992; Bell \& de Jong 2001; Kauffmann et al.\ 2003;
Panter et al.\ 2004; Gallazzi et al.\ 2005; McLure et al.\ 2009; Mortlock
et al.\ 2011). The observed GSMF is well fit by a simple or double Schechter
function (Baldry et al.\ 2008, 2012; Bolzonella et al.\ 2010; Pozzetti et
al.\ 2010) down to low mass limits 
($\mathcal{M} \sim 10^{8} \mathcal{M}_{\odot}$), and its analysis can be 
separated into more massive galaxies, 
$\mathcal{M}_{*} \gtrsim 10^{11} \mathcal{M}_{\odot}$, and less massive
ones. Although the GSMF is a well-established tool to study galaxy
evolution, our aim here is to develop a methodology capable of estimating
the GCMF using \textit{observational} data, since the GCMF itself is a
derived quantity and is therefore directly linked to the underlying
cosmological theory. The study of this function could provide
some insights about how, or even \textit{if}, effects of
relativistic nature can affect the mass evolution analysis.

The GCMF can be seen as an application of the general model connecting
cosmology theory to the astronomical data, introduced by RS03, and further
developed by Albani et al.\ (2007; hereafter A07) and Iribarrem et al.\
(2012; hereafter Ir12). These authors aimed at providing a relativistic
connection for the observed number counts data produced by observers and
studying its relativistic dynamics. In this paper we extend both goals
to the mass function of galaxies. This theoretical connection allows us
to study these quantities in other spacetime geometries than 
Friedmann-Lema\^itre-Robertson-Walker (FLRW), as we intend to do in
the future, and then trying to ascertain to what extent the underlying
choice of spacetime geometry affects these quantities, that is, to what
extent galaxy evolution might be affected by the spacetime geometry. 
Here we analyze the mass function for the average galactic mass at some
redshift interval and provide an illustration of our methodology by means
of deep galaxy redshift survey data. We use available observations of 
the galaxy luminosity function (LF), luminosity density, and stellar masses
to estimate the redshift evolution of the average galactic mass and 
luminosity. These two pieces of information are crucial to our analysis
because they cannot be obtained through cosmological principles, but 
have direct implications for a range of theoretical considerations
and the determination of an important quantity, the differential number
counts of galaxies. With the redshift evolution of $\mg$ and the 
equations presented in RS03 we can obtain the GCMF.

We used the LF parameters of the FORS Deep Field (FDF) galaxy survey
presented by Gabasch et al.\ (2004; hereafter G04) in the B band and
redshift range $0.5 < z < 5.0$ to calculate the selection function
$\psi$ and the luminosity density $j$, to then obtain the average
luminosity evolution, $L_{B}$. Next we computed the galaxy stellar
mass-to-light ratio using the galaxy stellar masses presented by Drory
et al.\ (2005). These two results lead to a redshift evolution of the
average galactic mass. Alternatively, we estimated $\mg(z)$
using the ratio between the stellar mass density and number density,
both quantities derived from the GSMF presented by Drory \& Alvarez
(2008) for the same FDF sample of galaxies. A comparison between these
two methodologies to calculate $\mg$ enables us to discuss the
intrinsic biases introduced by $\psi$, $j$ and the mass-to-light ratio
in the calculation of the average galactic mass. The next step was to
calculate $\drm \mg /\drm z$ and the theoretical quantities of interest,
such as the cumulative number counts $N$ and the differential number
counts $\drm N/\drm z$. Finally, the GCMF was computed assuming a
comoving volume, which allowed us to compare its results with predictions
from galaxy evolution models found in the literature.

The plan of the paper is as follows: \S 2 presents the relevant 
theoretical concepts to the analysis, 
\S 3 discusses the LF, selection function, and luminosity
density, and the general features of the FDF survey. 
\S 4 summarizes the semi-empirical relativistic approach
to obtain a GCMF, along with our results, and \S 5 presents our
conclusions. Here we adopted an FLRW cosmology with the following 
parameter values: $\Omega_{m_{0}} = 0.3$, $\Omega_{\Lambda_{0}}=0.7$, 
$H_{0} = 70 \ \textrm{km s}^{-1}\textrm{Mpc}^{-1}$.

\section{Theoretical framework}

We begin by assuming the spherically symmetric FLRW cosmology given
by the following line element,
\begin{equation}
\drm s^{2} = - c^{2} \drm t^{2} + S^{2} \left[\frac{\drm r^{2}}{1-kr^{2}}
+ r^{2}(\drm \theta^{2} + \sin^{2}\theta \drm \phi^{2}) \right],
\label{metric}
\end{equation}
where $t$ is the time coordinate and $r, \theta, \phi$ are the spatial 
coordinates, $S = S(t)$ is the cosmic scale factor, $k$ is the
curvature parameter ($k = +1, 0, -1$), and $c$ is the light speed.
The Einstein field equation with this line element yields the Friedmann
equation, which, if the cosmological constant $\Lambda$ is included, may be
written (e.g. Roos 1994)
\begin{equation}
H^{2} = \frac{8\pi G \rho_{m}}{3} + \frac{\Lambda}{3} - \frac{kc^{2}}{S^{2}},
\label{eq.friedmann}
\end{equation}
where $G$ is the gravitational constant,
$\rho_{m}$ is the matter density, and the Hubble parameter can be
defined as
\begin{equation}
H(t) \equiv \frac{1}{S(t)}\frac{\drm S(t)}{\drm t}, \quad \Longrightarrow 
\quad 
H_{0} = \frac{1}{S_{0}}\frac{\drm S_{0}}{\drm t},
\label{def.scale-factor}
\end{equation}
in which $S_{0}$ is the scale factor at the present time, and $H_{0}$ is
the Hubble constant. Note that the index zero is used to indicate 
quantities at the present time. 
Following Ir12, who used the law of conservation of energy in the
matter-dominated era and the past radial null geodesic, we find a
first-order ordinary differential equation for the scale factor in
terms of the radial coordinate $r$,
\begin{equation}
\frac{\drm S}{\drm r} = - H_{0} \left[
\frac{(\Omega_{\Lambda_{0}})S^{4} - S_{0}^{2}(\Omega_{0}-1)S^{2} + 
(\Omega_{m_{0}}S_{0}^{3})S}{c^{2} - H_{0}^{2}S_{0}^{2}(\Omega_{0}-1)r^{2}}
\right]^{1/2},
\label{eq.dSdr}
\end{equation}
where 
\begin{equation}
\Omega_{0} \equiv \Omega_{m_{0}} + \Omega_{\Lambda_{0}} =
\frac{\rho_{0}}{\rho_{0,c}} = \frac{\rho_{m_{0}}}{\rho_{0,c}} + 
\frac{\rho_{\Lambda_{0}}}{\rho_{0,c}},
\label{eq.dens-parameter}
\end{equation}
in which the critical density at the present time given by
\begin{equation}
\rho_{0,c} \equiv \frac{3H_{0}^{2}}{8\pi G},
\end{equation}
and the vacuum energy density in terms of the cosmological constant is
\begin{equation}
\rho_{\Lambda} \equiv \frac{\Lambda}{8\pi G}.
\end{equation}
Notice that since $\Lambda$ is a constant, then 
$\rho_{\Lambda} = \rho_{\Lambda_{0}}$.
To find numerical solutions for $S(r)$ we adopt $S_{0}=1$ and the 
cosmological parameters $\Omega_{m_{0}}=0.3$, $\Omega_{\Lambda_{0}}=0.7$
and $H_{0}=70$km s$^{-1}$Mpc$^{-1}$. We also used the fourth-order
Runge-Kutta method with the initial condition $r_{0}$ set to zero.

\subsection{Distances and volumes}
The area distance $d_{\ssty A}$, also known as angular diameter
distance, is defined by a relation between the intrinsically 
measured cross-sectional area element $\drm \sigma$ of the source and
the observed solid angle $\drm \Omega$ (Ellis 1971, 2007; Pleb\'anski
\& Krasi\'nski 2006), 
\begin{equation}
(d_{\ssty A})^{2} = \frac{\drm \sigma}{\drm \Omega} = 
\frac{S^{2}r^{2}(\drm \theta^{2} + \sin^{2} \theta \drm \phi^{2})}
{(\drm \theta^{2} + \sin^{2} \theta \drm \phi^{2})} = (Sr)^{2}.
\label{eq.dA}
\end{equation}

The luminosity distance $d_{\ssty L}$, which is defined as a relation
between the observed flux and the intrinsic luminosity of a source,
can be easily obtained from $d_{\ssty A}$, Eq.\ (\ref{eq.dA}), 
by invoking the Etherington reciprocity law (Etherington 1933;
Ellis 1971, 2007),
\begin{equation}
d_{\ssty L} = (1+z)^{2} d_{A},
\label{reciprocity-law}
\end{equation}
resulting in
\begin{equation}
d_{\ssty L} = S_{0}^{2}\left(\frac{r}{S}\right).
\label{eq.dL}
\end{equation}

The observations are usually obtained assuming a comoving volume 
$V_{\ssty C}$, but the theory often assumes a proper volume
$V_{\ssty Pr}$. From metric (\ref{metric}) the conversion of 
volume units can be given by
\begin{equation}
dV_{\ssty Pr} = \frac{S^{3}}{\sqrt{1-kr^{2}}} r^{2} \drm r \sin{\theta} 
\drm \theta \drm \phi = S^{3} \drm V_{\ssty C}.
\end{equation}
Hence, the relation between $n_{\ssty C}$ and $n$, which are the 
number densities of cosmological sources respectively given in terms
of comoving volume and proper volume, can be written as
\begin{equation}
n_{\ssty C} = S^{3} \ n.
\label{nc-n}
\end{equation}

\subsection{Differential number counts}
The general, cosmological model-independent, relativistic
expression for the number counts of cosmological sources $dN$ in a volume
section at a point down the null cone, and considering that both source
and observer are comoving, is given by Ellis (1971). This general 
expression was specialized to the FLRW cosmology by Ir12, yielding the
following expression:
\begin{equation}
\drm N = (d_{\ssty A})^{2} \drm \Omega \, n \, \frac{S}{\sqrt{1-kr^{2}}} \, \drm r,
\label{def.dN}
\end{equation}
where $n$ is the number density of sources per unit of proper volume in
a section of a bundle of light rays converging toward the observer and
subtending a solid angle $\drm \Omega$ at the observer's position. It 
can be related to the matter density $\rho_{m}$ and the average galactic
mass $\mg$ by means of
\begin{equation}
n = \frac{\rho_{m}}{\mg}.
\label{eq.n-rho}
\end{equation}
If we use the law of conservation of energy applied to the zero
pressure in the matter-dominated era, Eq. (\ref{eq.n-rho}) becomes
\begin{equation}
n = \left(\frac{3\Omega_{m_{0}} H_{0}^{2}S_{0}^{3}}{8\pi G \mg} \right)
\frac{1}{S^{3}}.
\label{eq.n}
\end{equation}
Considering $\drm \Omega=4\pi$, Eqs.\ (\ref{eq.dA}) and (\ref{eq.n}), 
we can rewrite Eq. (\ref{eq.friedmann}) at present time as follows: 
\begin{equation}
kc^{2} = H_{0}^{2} S_{0}^{2} (\Omega_{0}-1),
\label{eq.kc2}
\end{equation}
and Eq.\ (\ref{def.dN}) as
\begin{equation}
\frac{\drm N}{\drm r} = \left(\frac{3\ c \Omega_{m_{0}} H_{0}^{2} 
S_{0}^{3}}{2G\mg} \right)
\left[\frac{r^{2}}{\sqrt{c^{2}-H_{0}^{2}S_{0}^{2} (\Omega_{0} - 1)
r^{2}}} \right].
\label{eq.dNdr}
\end{equation}

The redshift $z$ can be written as
\begin{equation}
1+z = \frac{S_{0}}{S},
\label{def.z}
\end{equation}
where a numerical solution of the scale factor immediately gives us a
numerical solution for $z(r)$. In this way, we can obtain the differential
number counts $dN/dz$ by means of the following expression:
\begin{equation}
\frac{\drm N}{\drm z} = \frac{\drm N}{\drm r} \frac{\drm r}{\drm S} 
\frac{\drm S}{\drm z}.
\label{def.dNdz}
\end{equation}
These derivatives can be taken from Eqs. (\ref{eq.dSdr}),
(\ref{eq.dNdr}), and (\ref{def.z}), enabling us to write
\begin{eqnarray}
\frac{dN}{dz} = & & \left( \frac{3 \; c \; \Omega_{m_0} H_0 {S_0}^2}{2 G
                \mg} \right) \times \nonumber  \\ 
  & & \times \left[ \frac{r^2 S^2}{ \sqrt{ 
		(\Omega_{\Lambda_0})S^4 - {S_0}^2
		(\Omega_0-1)S^2 + (\Omega_{m_0}{S_0}^3)S}} \right].
\label{dNdz2}
\end{eqnarray}

\section{Galaxy luminosity function and stellar mass function}
The galaxy LF $\phi(L, z)$ gives the number density of galaxies with
luminosity $L$ at redshift $z$. In the Schechter (1976) analytical
form it is written as,
\begin{equation}
\mathnormal{\phi(L) \, \drm L =\frac{\phi^{*}}{L^{*}} \ 
\left(\frac{L}{L^{*}}\right)^{\alpha} \ \exp{\left(-\frac{L}{L^{*}}\right)}
\, \drm L = \phi(\ell) \, \drm \ell },
\end{equation}
where $\mathnormal{\ell\equiv L/L^{*}}$, $\mathnormal L$ is the observed
luminosity, $\mathnormal L^{*}$ is the luminosity scale parameter,
$\mathnormal \phi^{*}$ is the normalization parameter, and $\mathnormal
\alpha$ is the faint-end slope parameter. These parameters are
determined by careful analysis of data from galaxy redshift surveys. 
The selection function $\psi$ in a given waveband above the lower 
luminosity threshold $\ell_{lim}$ is written as
\begin{equation}
\mathnormal \psi(z) = \int_{\mathnormal \ell_{lim}(z)}^{\infty}
\mathnormal{\phi(\ell)  \ \drm \ell},
\label{eq.psi}
\end{equation}
where 
\begin{equation}
\ell_{lim}(z) = \frac{\mathnormal L_{lim}}{\mathnormal L^{*}} = 
10^{0.4 (M^{*} - M_{lim})},
\end{equation}
\begin{equation}
M_{lim}(z) = m_{lim} - 5\log[d_{L}(z)] - 25 + A^{l}.
\end{equation}
Here $M^{*}$ is the absolute magnitude scale parameter (it is 
connected to $L^{*}$), $d_{L}$ is the luminosity distance, $m_{lim}$
is the limiting apparent magnitude of the survey, and $A^{l}$ is the
reddening correction.
 
The luminosity density $j(z)$ provides an estimate of the total
amount of light emitted by galaxies per unit volume in a given band. It
can be obtained from the following integral of the observed LF in a given
band:
\begin{equation}
j(z) = \int_{\ell_{lim}(z)}^{\infty} \ell \ \phi(\ell) \ \drm \ell = L^{*} \ \phi^{*}
\ \Gamma\left(\alpha + 2, \frac{L_{lim}}{L^{*}}\right),
\label{eq.j}
\end{equation}
where $\Gamma(a,x)$ is the incomplete gamma integral such that
$\lim\limits_{x \rightarrow 0} \ \Gamma(a,x) = \Gamma(a)$.

Similarly to the LF, the GSMF can be written in a Schechter form,
\begin{equation}
\bar{\phi}(\mathcal{M}) = \ln(10) \, \bar{\phi^{*}}\left[10^{(\mathcal{M}-
\mathcal{M}^{*})(1+\bar{\alpha})}\right]\times \exp \left[-10^{(\mathcal{M}
-\mathcal{M}^{*})}\right],
\label{eq.GSMF-Schechter}
\end{equation}
where $\bar{\phi^{*}}$ is the GSMF normalization parameter, $\bar{\alpha}$
is the faint-end slope and $\mathcal{M}^{*}= \log(\mathcal{M}^{*}_{stellar}
/\mathcal{M}_{\odot})$ is the characteristic mass that separates the exponential
part of the function, dominant at high masses, and the power-law part, important
at low masses. The GSMF can also be represented by a double-Schechter function
that includes a second power-law. Here we only use the simple Schechter
function.

From the GSMF, two other quantities can be defined, the stellar
mass density,
\begin{equation}
\rho_{*}(z) = \int_{\mathcal{M}_{lim}}^{\infty} \mathcal{M} \ \bar{\phi}(\mathcal{M}) \, 
\, \drm \mathcal{M} = \mathcal{M}^{*} \ \bar{\phi^{*}}
\ \Gamma\left(\bar{\alpha} + 2, \frac{\mathcal{M}_{lim}}{\mathcal{M}^{*}}\right),
\label{eq.rho-star}
\end{equation}
and the number density of galaxies 
\begin{equation}
n_{*}(z) = \int_{\mathcal{M}_{lim}}^{\infty} \bar{\phi}(\mathcal{M}) \, 
\drm \mathcal{M},
\label{eq.n-star}
\end{equation}
for galaxy stellar masses above a given $\mathcal{M}_{lim}$. 
This lower mass limit is uncertain and related to the magnitude
limit of the survey, which in itself depends on the redshift, just as the
lower luminosity $L_{lim}(z)$ was used to calculate the luminosity and 
number density from the LF. However, for our purposes here, to avoid a 
bias caused by selection effects introduced by the limits of the survey,
we derived the total number and mass densities extrapolating the
integrals above to the lower masses, independently of the redshift. The
choice of $\mathcal{M}_{lim}$ is addressed in the next sections.

\subsection{FORS Deep Field Galaxy Survey dataset}
The FDF is a multicolor photometric and spectroscopic survey of a
$7^{\, \prime} \times 7^{\, \prime}$ region near the south galactic pole.
According to G04, the sample is composed of 5558 galaxies selected in
the I band and photometrically measured to an apparent magnitude
limit of $I_{AB}=26.8$. The absolute magnitude of each galaxy in the
sample was computed by G04 using the best-fitting spectral energy
distribution (SED) given by the photometric redshift convolved with
the appropriate filter function and K-correction. The stellar
mass-to-light ratios 
$\mathcal{M}_{*}/L_{\ssty B}$ for the galaxies in the catalog were
estimated with a log-likelihood-based SED-fitting technique, using
a library of SEDs built with the stellar population evolution model
given by Bruzual \& Charlot (2003) and a Salpeter (1955) initial
mass function. These $\mathcal{M}_{*}/L_{\ssty B}$ data
were obtained via private communication with Niv Drory,
and are based on the analysis described in Drory et al.\ (2005).
Note that $\mathcal{M}_{*}/L$ at $z > 2.5$ might be 
overestimated because of less reliable information on the rest-frame
optical colors at young mean ages.

The LF parameters derived from this sample for the B band
by G04 are for the redshift range $0.5\leq z \leq 5.0$.
The authors also proposed the following equations for the redshift
evolution of these parameters:
\begin{eqnarray}
\phi^{\ast}(z) & = & \left(0.0082^{+0.0014}_{-0.0012}\right) \, 
(1+z)^{-1.27^{+0.16}_{-0.19}}, \\
M^{\ast}(z) & = & \left(-20.92^{+0.32}_{-0.25}\right) + 
\left(-1.03^{+0.23}_{-0.28}\right) \ln (1+z), \\
\alpha(z) & = & -1.25\pm0.03.
\end{eqnarray}

The GSMF for the same FDF galaxy sample was calculated by 
Drory et al. (2005) and further analyzed by Drory \& Alvarez (2008)
in the context of the contribution of star formation and merging to
galaxy evolution. These authors assumed the mass function to be
of a simple Schechter form. For the present work, we used the table
with the Schechter fit parameters from Drory \& Alvarez (2008), which
were obtained using the $1/V_{\mathrm{max}}$ method in seven bins
from $z = 0.25$ to $z=5.0$. It is interesting to note that the
faint-end slope $\bar{\alpha}$ is constrained to the redshift range
$0 < z < 2$, where the authors considered the data to be deep enough
and found it to be given by a constant, $\bar{\alpha}(z) = -1.3$.
Because the data do not allow $\bar{\alpha}$ to be constrained at
higher redshifts, this value of the faint-end slope is extrapolated
to $z > 2$.

\section{The galaxy cosmological mass function}
\subsection{Average galactic mass}
\begin{figure}[htb]
\centering
\includegraphics[width=0.49\textwidth]{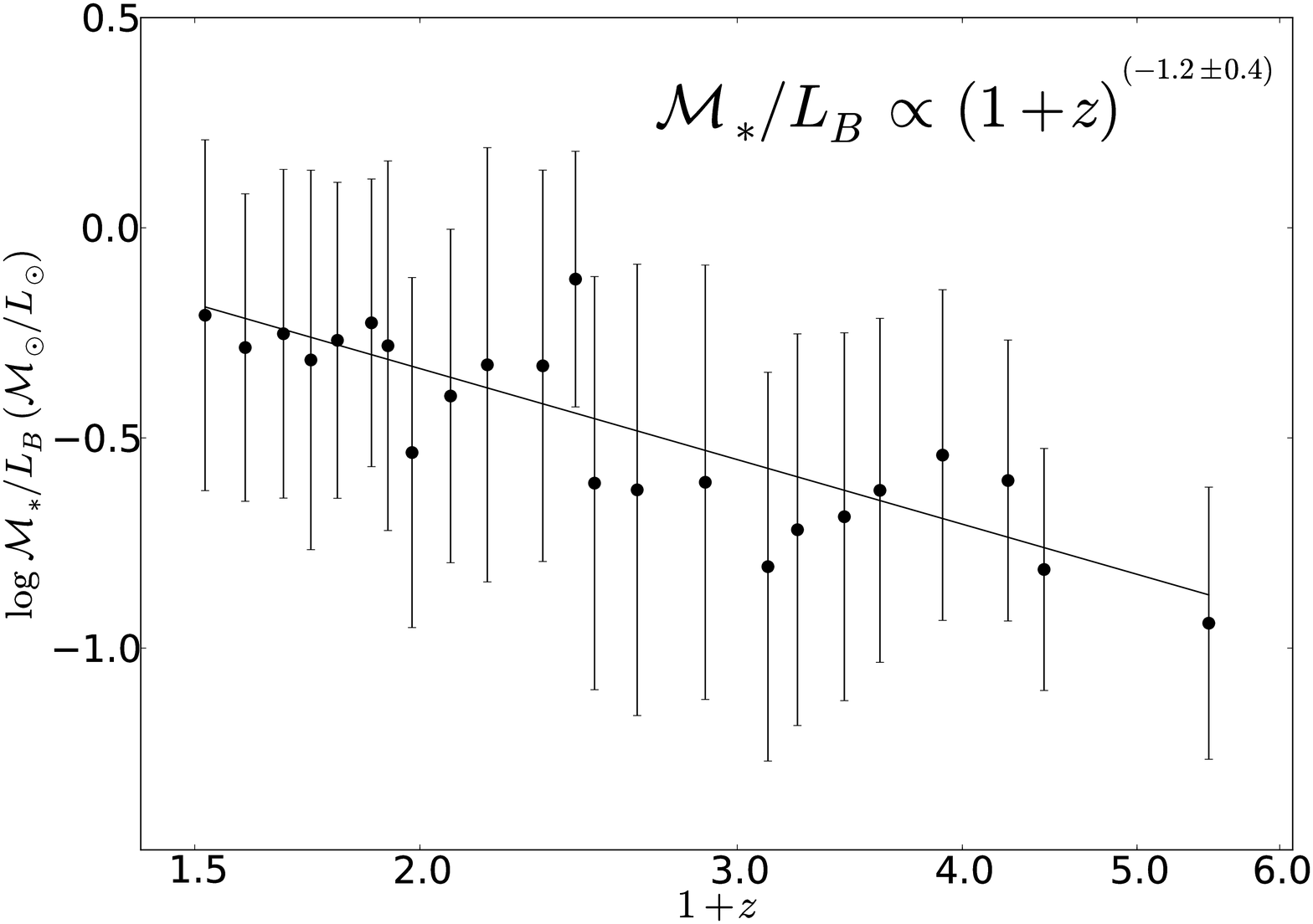}

\includegraphics[width=0.49\textwidth]{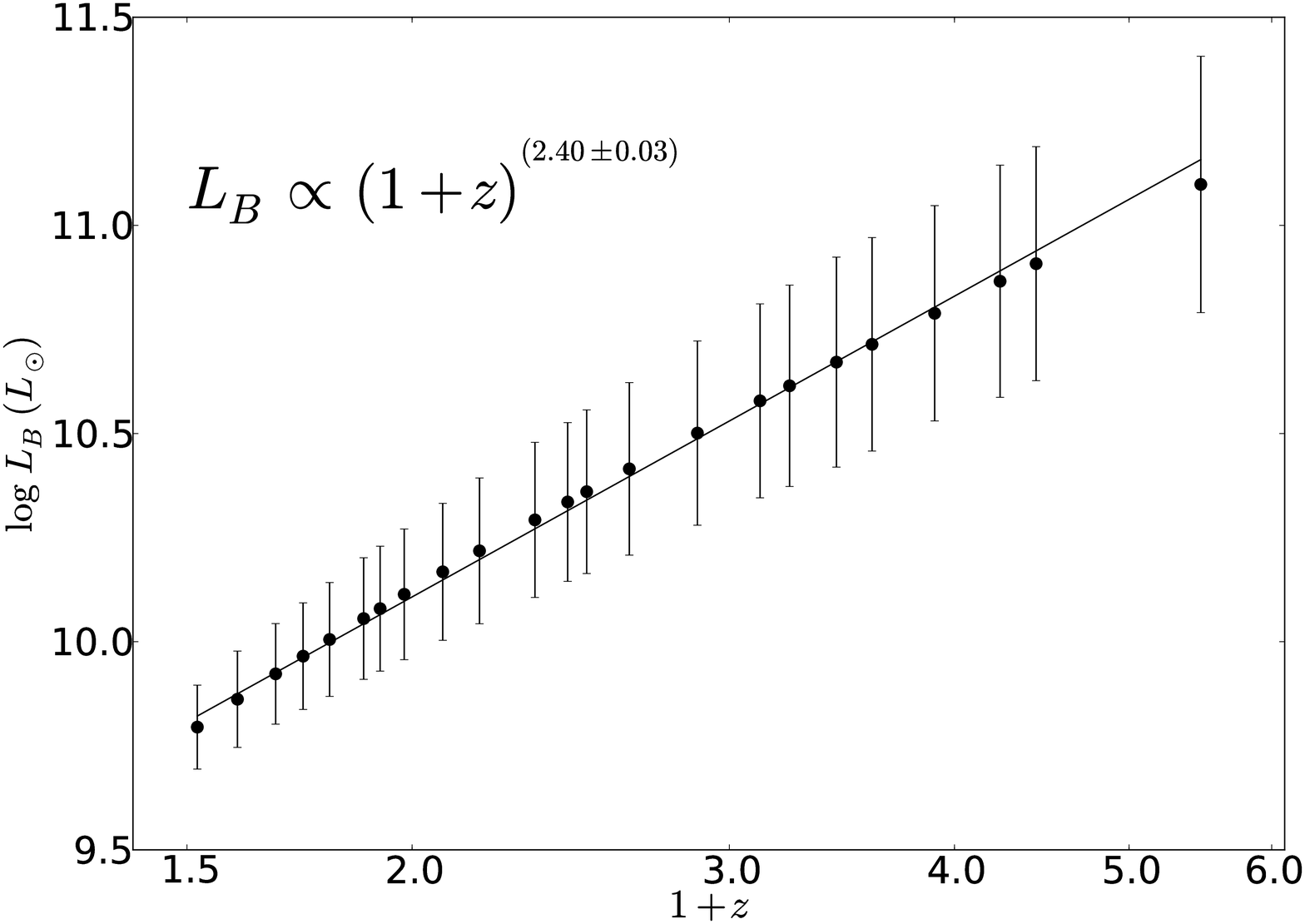}
\caption{\textit{Upper panel:} Redshift evolution of the galaxy
stellar mass-to-light
ratio in the B band for the FDF data. The graph shows a power-law fit
in terms of the redshift. \textit{Lower panel:} Redshift evolution of
the average galaxy luminosity of the FDF dataset in the B band and its
corresponding power-law data fit.}
\label{ml-l}
\end{figure}

To discuss the semi-empirical relativistic approach to estimate
the mass function, we start with the proposal of RS03 for an
expression of the mass-to-luminosity ratio at a given redshift value,
\begin{equation}
\frac{\mathcal{M}}{L} = \mg(z)\frac{\psi(z)}{j(z)}.
\label{def.ml}
\end{equation}
From the observational point of view, LF catalogs only give us
information about the stellar mass $\mathcal{M_*}$ and stellar
mass-to-light ratio $\mathcal{M_{*}}/L$ of the galaxies. 
But, assuming that $\mathcal{M_{*}}/L$ is proportional to $\mg/L$,
we can write the following expression,  
\begin{equation}
\frac{\mathcal{M_{*}}}{L_{\ssty B}} \propto 
\frac{\mathcal{M}}{L_{\ssty B}} \propto 
\mg(z)\frac{\psi_{\ssty B}(z)}{j_{\ssty B}(z)}.
\label{eq.ml}
\end{equation}
Since we have $\mathcal{M_{*}}/L_{\ssty B}$ for each galaxy
in the FDF sample, we can calculate the average galaxy
stellar mass-to-light ratio using a subsample of 201 galaxies
per redshift bin. Next, we can employ Eq.\ (\ref{def.ml}) to
estimate the average galaxy luminosity in a given passband,
\begin{equation}
L_{\ssty B} \propto \frac{j_{\ssty B}(z)}{\psi_{\ssty B}(z)},
\end{equation}
and use LF data from the FDF survey to ascertain the general
behavior of the average luminosity in terms of the redshift.
Fig.\ \ref{ml-l} shows the results of the average galaxy
stellar mass-to-light ratio and the average luminosity in the
B band. Both quantities can be fitted by power-law relations,
and the results are as follows:
\begin{eqnarray}
\mathcal{M}^{*}/L_{B} \propto (1+z)^{-1.2\pm0.4}, \\ 
L_{B} \propto (1+z)^{2.40 \pm 0.03}.
\end{eqnarray}
The error bars for $L_{B}$ were obtained by
Monte Carlo simulations. Hence, Eq.\ (\ref{eq.ml}) allows us
to estimate the average galactic mass from observations. The 
result is as follows:
\begin{equation}
\mg(z) \propto  \frac{\mathcal{M_{*}}}{L_{\ssty B}} 
\frac{j_{\ssty B}(z)}{\psi_{\ssty B}(z)}.
\label{eq.mass}
\end{equation}
This expression entails that in general the total galactic
mass follows its luminous mass evolution, that is, more dark
matter implies more stars when one considers galaxies as a
whole and not regions of galaxies, for example, extended dark
matter halos. We are aware that the previous assumption seems
to be reasonable for early-type galaxies, that is, for ellipticals
and lenticulars (e.g., Magain \& Chantry 2013), but it contrasts
with rotation curves from spirals. Because our data do not have
any morphological classification, the present approach is
suitable to our sample.

\begin{figure}[htb]
\centering
\includegraphics[width=0.49\textwidth]{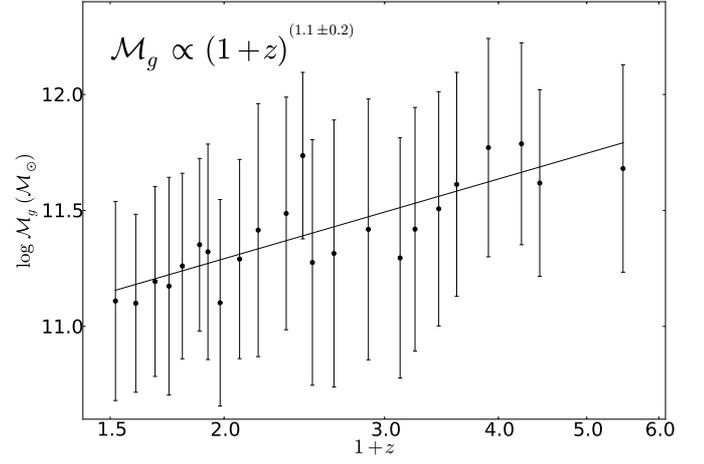}
\caption{Redshift evolution of the average galactic mass for the FDF data, 
using LF and mass-to-light ratio data.
As shown in the graph, the data points can be fitted by a mild power law.
The determination coefficient for this fit is $R^{2} = 0.64$, where
$R^{2}$ is a statistical measure of how well the regression line approximates
the real data points. It ranges from 0 to 1, and a value of $R^{2}=1$
indicates that the regression line perfectly fits the data.}
\label{mass-counts}
\end{figure}

We emphasize that the $\mg$ derived from 
$\mathcal{M}_{*}/L_{\ssty B}$ depends on the variation in the
spectral type of the galaxy: early-type galaxies have more tightly
constrained masses than late-type galaxies. This means that the
resulting $\mg$ may present a large dispersion due to the lack of
morphological classification in our sample. However, at high $z$
the uncertainty in $\mathcal{M}_{*}/L$ increases because objects
drop out in the blue bands and stellar populations
become younger. The behavior of $\mg$ can be seen in 
Fig.\ \ref{mass-counts}, in which the simplest description is as a single
power-law, given by
\begin{equation}
\mg = {\mathcal{M}_{g_0}} (1+z)^{1.1\pm0.2},
\label{result1}
\end{equation}
where ${\mathcal{M}_{g_0}} \approx 10^{11} \mathcal{M}_\odot$ is
the assumed local value of the average galactic mass (Sparke \&
Gallagher 2000). Nevertheless, other interpretations than a single
power-law are possible because of the large dispersion in $\mg$. 

\begin{figure}[htb]
\centering
\includegraphics[width=0.49\textwidth]{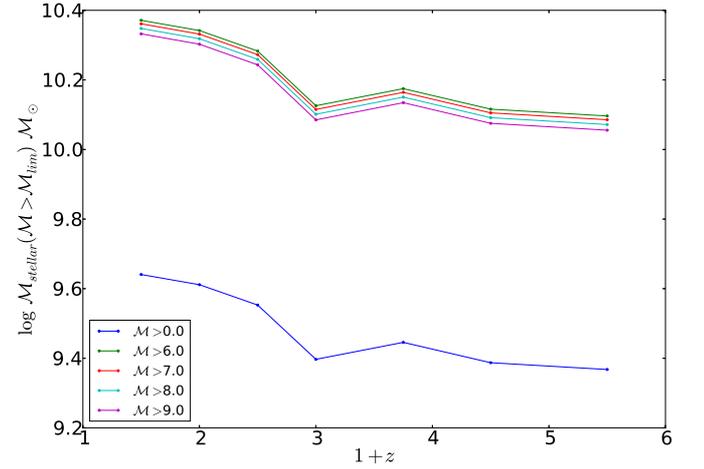}
\caption{Average stellar mass estimated using the GSMF data from
FDF sample for different lower mass limits. In this graph the error
bars were omitted to emphasize the behavior of $\log{\mathcal{M}_{stellar}}$
along the redshift range with different mass limits.}
\label{Mass-Star}
\end{figure}

\begin{figure}[htb]
\centering
\includegraphics[width=0.49\textwidth]{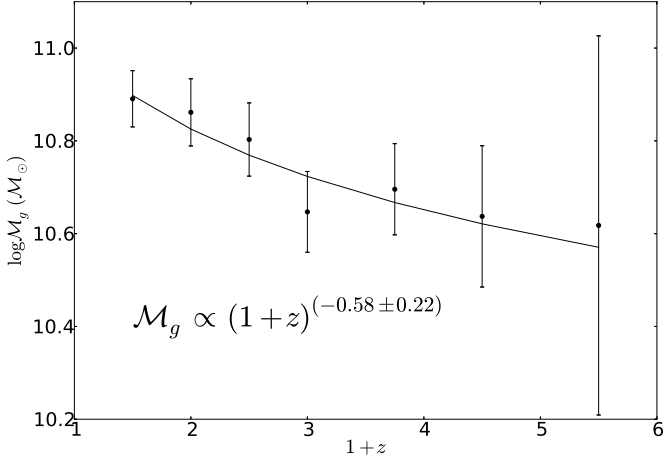}
\caption{Redshift evolution of the average galactic mass for the FDF survey
using GSMF data. The plot shows that the data points can be fitted by 
a mild power-law decrease. The determination coefficient for this fit
is $R^{2} = 0.84$.}
\label{mass-mf}
\end{figure}

The average galactic mass can also be derived by following
the alternative approach of using quantities derived from the GSMF,
$\rho_{*}$ and $n_{*}$, which yield the average galactic stellar mass
$\mathcal{M}_{stellar}$.\footnote{We are grateful to the referee for 
suggesting this procedure.} Still under the assumption that the total
galactic mass evolves as the luminous mass, one may write the
following expression, 
\begin{equation}
\mg(z) \propto \mathcal{M}_{stellar}(z) \propto \frac{\rho_{*}}{n_{*}}.
\end{equation}
As these quantities depend on the lower mass limit $\mathcal{M}_{lim}$
used in the integration of Eqs.\ (\ref{eq.rho-star}) and
(\ref{eq.n-star}), we tested different values for $\mathcal{M}_{lim}$
to check the possible influence of this limit in our results. The
results are presented in Fig.\ \ref{Mass-Star}
and clearly show that the various values for $\mathcal{M}_{lim}$
will only affect the amplitude of $\mathcal{M}_{stellar}$, but not
its general behavior. Our goal is to obtain a general form for the
average galactic mass, hence assuming that the mass evolves as a
power-law we can freely choose a lower mass limit in our calculations.
We adopted $\mathcal{M}_{lim} \approx 0$ and $\mathcal{M}_{g_0}
\approx 10^{11} \mathcal{M}_{\odot}$ to carry out our power-law
fitting. The results are shown in Fig.\ \ref{mass-mf} and the fitted
expression is written as
\begin{equation}
\mg = {\mathcal{M}_{g_0}} (1+z)^{-0.58\pm0.22}.
\label{eq.mg-result}
\end{equation}

Comparing this expression with Eq.\ (\ref{result1}) shows
that the two methodologies discussed produce different results when 
they are employed to calculate the average galactic mass in terms of the
redshift. This is due to the survey limits, since the former method
explicitly depends on these limits, as can be seen in the definition of
the selection function and the luminosity density, respectively given
by Eqs.\ (\ref{eq.psi}) and (\ref{eq.j}). Moreover, the average
mass-to-light ratio can also introduce a bias because one cannot
distinguish the changes in $\mathcal{M}_{*}/L$ as being due to either real
changes of the stellar population with redshift or due to the fact that
brighter objects are selected as having different $\mathcal{M}_{*}/L$ values.
The stellar mass data were obtained with the latter metthod, using the
SED fitting, which used a large range of wavelength observations,
and therefore produced less biased results than the former. This is
expected to make the second method more reliable, and we therefore
use Eq.\ (\ref{eq.mg-result}) from this point on.

The negative power index in Eq.\ (\ref{eq.mg-result})
indicates a growth in mass from high to low redshift values. Indeed,
from this equation we can see that galaxies at $z=5$ had on average from
25\% to 50\% of their present masses at $z=0$. This growth might be caused 
by the galaxy mergers within the FDF redshift range, by the star formation
history itself, or even by a combination of these two effects. However,
one cannot separate these two mechanisms using stellar mass
data.

Finally, we stress that Fig.\ \ref{mass-mf} shows the
redshift evolution of $\mg$ based on the FDF survey. Nevertheless,
results obtained by means of data from a single survey are uncertain
and, therefore, a more robust estimate requires data from different
surveys. We intent to do so in forthcoming papers.

\subsection{Observational quantities}
The differential number counts in the expression (\ref{dNdz2}) is 
directly linked to the underlying cosmological model, since this is a
theoretical quantity given by relativistic cosmology. Therefore, we
need to write $\drm N/\drm z$ in terms of observational quantities,
which can be achieved by using the methodology developed by RS03, and
further extended in A07 and Ir12, which connects this theoretical
quantity to the LF. The link between relativistic cosmology theory
and observationally determined LF is achieved by using the
\textit{consistency function} $J(z)$, which represents 
the undetected fraction of galaxy counts in relation to the one
predicted by theory as follows:
\begin{equation}
[\drm N]_{\mathrm{obs}} = J(z) \ \drm N.
\label{eq.J-dN}
\end{equation}
Here the observed differential number counts $[\drm N]_{\mathrm{obs}}$
is the key quantity for our analysis because other quantities require 
its previous knowledge. 

From expressions (\ref{nc-n}) and (\ref{def.dN}), we obtain
\begin{equation}
dN = (d_{\ssty A})^{2} \drm \Omega \left(\frac{n_{\ssty C}}{S^{3}}\right)
\frac{S}{\sqrt{1-kr^{2}}} \drm r.
\label{dN}
\end{equation}
Then, to derive $[\drm N]_{\mathrm{obs}}$ we need the observational 
counterpart of $n_{\ssty C}$, which is, according to its definition, the 
selection function $\psi$. Therefore,
\begin{equation}
[\drm N]_{\mathrm{obs}} = (d_{\ssty A})^{2} \drm \Omega 
\left(\frac{\psi}{S^{3}}\right) \frac{S}{\sqrt{1-kr^{2}}} \drm r.
\label{dNobs}
\end{equation}
The substitution of Eqs. (\ref{dN}) and (\ref{dNobs}) into 
Eq. (\ref{eq.J-dN}) yields 
\begin{equation}
\psi(z) = J(z)\ n_{\ssty C}.
\label{def.J}
\end{equation}

For our purposes it is more convenient to express the number counts 
$dN$ in terms of the redshift,
\begin{equation}
\left[\frac{\drm N}{\drm z} \right]_{\mathrm{obs}} = J(z) 
\frac{\drm N}{\drm z},
\label{eq.dNdzobs-J}
\end{equation}
and considering Eq. (\ref{def.J}), it can be rewritten as
\begin{equation}
\left[\frac{\drm N}{\drm z} \right]_{\mathrm{obs}} = 
\frac{\psi}{n_{\ssty C}} \frac{\drm N}{\drm z},\quad \Rightarrow \quad 
\left[\frac{\drm N}{\drm z} \right]_{\mathrm{obs}} = 
\frac{V_{\ssty C}}{V_{\ssty Pr}} \frac{\psi}{n} \frac{\drm N}{\drm z},
\label{eq.dNdz-obs}
\end{equation}
where the two volume definitions appear in the expression above because
the relativistic number counts are originally defined in a proper
volume, therefore one requires a suitable volume transformation. $V_{\ssty C}$,
$V_{\ssty Pr}$, $n_{\ssty C}$, $n$ and $\drm N /\drm z$ are theoretical
quantities obtained from the underlying spacetime geometry and, hence,
they need to be determined in the chosen cosmological model so that we
can obtain the observational differential number counts of 
Eq.\ (\ref{eq.dNdz-obs}). The only nontheoretical quantity in this 
equation is the selection function. In addition, as already
discussed in A07 and Ir12, if we substitute Eqs.\ (\ref{eq.n}) and
(\ref{dNdz2}) into Eq.\ (\ref{eq.dNdz-obs}), the term $\mg$ cancels
out and renders $\obs{\drm N / \drm z}$ mass independent on first order.

\begin{figure}[htb]
\centering
\includegraphics[width=0.49\textwidth]{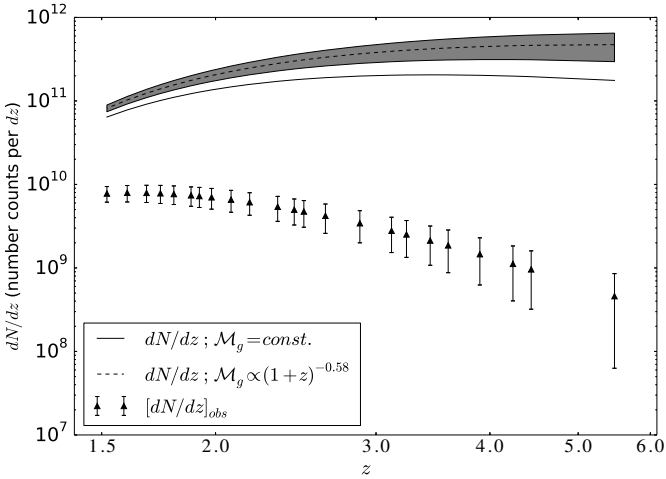}
\caption{Redshift evolution of the theoretical differential
number counts using constant and evolving values for
$\mathcal{M}_{g}$, as well as the observational values. The constant
value used was the assumed local ($z \approx 0$) average galactic
mass $\mathcal{M}_{g} = 10^{11} \mathcal{M}_{\odot}$. Symbols
are as in the legend, and the gray area is the $1\sigma$ error of 
the power index of $\mg(z)$.}
\label{fig.dNdz}
\end{figure}

Now, considering Eq. (\ref{eq.mg-result}) we assume two
possible cases for the average galactic mass: $\mg \approx 10^{11}
\mathcal{M}_{\odot}$ for all redshift ranges and $\mg \propto
(1+z)^{-0.58\pm 0.22}$. Substituting both cases in Eq. (\ref{dNdz2}),
we can see the implication on the differential number counts 
of an evolving average galactic mass. Fig.\ \ref{fig.dNdz} shows the
behavior of the theoretical differential number counts $\drm N/\drm z$
using a constant and evolving $\mg$, as well the values of $\obs{\drm
N/ \drm z}$. The change from constant to evolving $\mg$ does not
affect the general behavior of $\drm N/\drm z$ in a significant way.
Therefore, assuming a constant value for $\mg$, as was done in RS03,
A07, and Ir12, can be considered as a very reasonable analytical
simplification of the problem and, hence, the conclusions reached
by these authors hold in general, at least as far as the FDF survey
is concerned.

\subsection{The cosmological mass function of galaxies}
As stated above, the GCMF contains information about the
galactic number density at a certain redshift in terms of the average
galactic mass $\mg(z)$. It can be defined as follows:
\begin{equation}
\zeta[\mg(z), z] \equiv \frac{1}{V_{\ssty C}} \frac{\drm N}{\drm 
     \left(\log \mg \right)} = \frac{1}{V_{\ssty C}} { \left[ \frac{\drm
     \left( \log \mg \right) }{\drm z}\right] }^{-1}
\frac{\drm N}{\drm z}.
\label{eq.MF}
\end{equation}
Here we followed the standard practice in GSMF calculations of
writing the galaxy mass function in terms of logarithmic mass and the
comoving volume in which the GCMF is derived, since it is now standard
practice to calculate $\phi(L,z)$ in terms of $V_{\ssty C}$.

Substituting Eq.\ (\ref{eq.dNdzobs-J}) into Eq. (\ref{eq.MF}), we can
write the following expression
\begin{equation}
\zeta(z) = \frac{1}{V_{\ssty C}} { \left[ \frac{\drm
     \left( \log \mg \right) }{\drm z}\right] }^{-1}
\frac{1}{J(z)}\left[\frac{\drm N}{\drm z} \right]_{\mathrm{obs}}.
\end{equation}
Then, we can also define
\begin{equation}
[\zeta]_{\mathrm{obs}}(z) \equiv 
\zeta (z) \, J (z) = \frac{1}{V_{\ssty C}}
\frac{[\drm N/\drm z]_{\mathrm{obs}}}{\drm \left( \log \mg \right) /\drm z}.
\end{equation}

\begin{figure}
\centering
\includegraphics[width=0.49\textwidth]{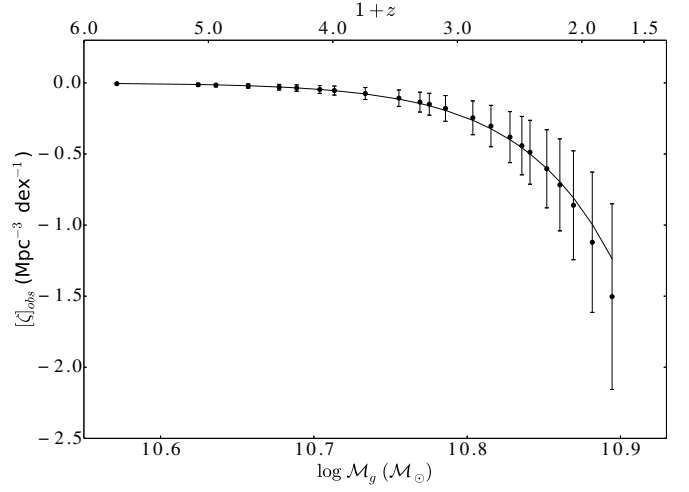}
\caption{Galaxy cosmological mass function 
in terms of the average galactic mass and its corresponding redshift
evolution. The best-fit function has $\chi^{2} =0.029$. The function
is given by a Schechter type function (Eq.\ \ref{eq.GSMF-Schechter}) and
the fitted parameters
are given in Eqs.\ (\ref{lf1}), (\ref{lf2}) and (\ref{lf3}).}
\label{zeta}
\end{figure}

Fig.\ \ref{zeta} shows the relationship between the GCMF and $\mg$, as
well as its redshift dependence. We note that the GCMF is negative, which
is not caused by a logarithmic effect, but a consequence
of the method used to infer $\drm ( \log \mg ) / \drm z$. Thus, the number
density of galaxies whose masses lie in the range $\mg$, $\mg + \drm \mg$
at the redshift range $z$, $z + \drm z$ is given by 
$\zeta \, \drm \mg$, and not simply by the function $\zeta$.

The GCMF $\obs{\zeta}(1+z)$ vs.\ $\log \mg$ data can be fitted by a
Schechter function of the form given by Eq.\ (\ref{eq.GSMF-Schechter}). The
best-fit parameters are 
\begin{eqnarray}
\bar{\phi^{*}} & = & -0.2 \pm 0.5 \; \textrm{Mpc}^{-3}, \label{lf1} \\
\log \mathcal{M}^{*} & = & 10.8 \pm 0.1 \; \mathcal{M}_{\odot}, \label{lf2} \\
\bar{\alpha} & = & 7.5 \pm 0.7. \label{lf3}
\end{eqnarray}
Although functionally similar to the GSMF, $[\zeta]_{\mathrm{obs}}$
was derived using a different methodology and, therefore, it is not
directly comparable with the GSMF mass function found in the literature. Hence,
the Schechter parameters from both functions are, in principle, unrelated.
In addition, a direct comparison with the GSMF (e.g., Drory
et al.\ 2004, 2005; Bundy et al.\ 2006; Pozzetti et al.\ 2007) is not
possible because we analyze the average mass where we cannot see this
differential behavior for the different bins of mass, this being a standard
approach used to study the galaxy stellar masses. However, we intend to
extend our analysis to use different bins of mass and include the study of
the barionic matter evolution instead of only using the stellar mass.

The result obtained for the GCMF suggests that on average galaxies
were less massive in the past than in the present, a behavior that agrees
with predictions from the ``bottom-up" (small objects form first) assembly of
dark matter structures in cold dark matter models. We also note that
there is a strong variation on the GCMF in the range $0.5 < z < 2.0$, which
can be interpreted as being a result of galaxy mergers or the evolution
of the galaxy star formation history itself, as mentioned above.

As last remarks, we recall the limitations of the
sample used and that the lack of morphological classification might
imply that two or more different types of galaxies may cause
different effects in the GCMF. Therefore, more analyses with different
datasets need to be made.

\section{Conclusion}

We discussed a semi-empirical relativistic
approach capable of calculating the observational galaxy
cosmological mass function (GCMF) in a relativistic cosmology
framework. The methodology consists of employing the luminosity
function results obtained by G04 using the B-band FORS Deep Field
galaxy survey data in the redshift range $0.5 < z < 5.0$ to
calculate the selection function and luminosity density, which led us to
conclude that the galaxy average luminosity in this
sample behaves according to $L_{\ssty B} \propto (1+z)^{(2.40\pm0.03)}$.
From the stellar mass-to-light ratio of the same galaxy sample, we found
that on average this quantity presents a power-law behavior,
$\mathcal{M}_{*}/L_{\ssty B} \propto (1+z)^{(-1.2\pm0.4)}$. These 
results led to a redshift evolution of the average galactic mass
given by $\mg \propto (1+z)^{(1.1\pm0.2)}$, that is, a power-law behavior
with positive power index.

Alternatively, $\mg(z)$ was also estimated by means of the galaxy
stellar mass function (GSMF) data, which resulted in a power law
with a negative power index, given by $\mg \propto (1+z)^{(-0.58 \pm 0.22)}$.
We found the former approach less reliable because of its strong
dependence on the selection function and luminosity density with the
limit of the survey. This produced more strongly biased results, and
we adopted the latter result in our calculations.

We then derived the theoretical quantities using the technique
discussed in RS03, A07, and Ir12, which enabled us to compute the
observational GCMF in the FLRW metric. We found that the GCMF decreases
as the galactic average mass increases, this pattern is well
fitted by a Schechter function with very different parameters values from
the values found in literature for the GSMF. This general behavior seems
to support the prediction of cold dark matter models in which the less
massive objects are formed earlier. Moreover, in the
range of $0.5 < z < 2.0$ the GCMF varies strongly, which might
be interpreted to be a result of a high number of galaxy
mergers in more recent epochs or as a strong evolution in the star
formation history of these galaxies.

\begin{acknowledgements}
We are grateful to N.\ Drory for kindly providing the necessary data
to develop this work. A.R.L.\ and A.I.\ respectively acknowledge the
financial support from the Brazilian agencies FAPERJ and CAPES. We are
also grateful to very helpful suggestions made by the referee.
\end{acknowledgements}

\vspace{2cm}

\end{document}